\documentclass[aps, prd, twocolumn, nofootinbib,]{revtex4-2}

\usepackage{amsmath}
\usepackage{amsfonts}
\usepackage{wasysym}
\usepackage{graphicx}
\usepackage{comment}
\usepackage{tensor}

\def\filetype{pdf}
\def\path{}

\allowdisplaybreaks[1]

\begin{document}



\title{Dynamical evolution of fermion-boson stars with realistic equations of state}
\author{Joseph E.\ Nyhan and Ben Kain}
\affiliation{Department of Physics, College of the Holy Cross, Worcester, Massachusetts 01610, USA}

\begin{abstract}
\noindent Fermion-boson stars are mixtures of the ordinary nuclear matter of a neutron star and bosonic dark matter.  We dynamically evolve fermion-boson stars for the first time using a realistic equation of state for nuclear matter.  We use our dynamical solutions to make a detailed study of the evolution of weakly and strongly perturbed static solutions.  As examples of our findings, we identify a region of parameter space where weakly perturbed unstable static solutions migrate to a stable configuration and we determine the criteria under which strongly perturbed stable static solutions will always move to a stable configuration instead of collapsing to a black hole.
\end{abstract} 

\maketitle


\section{Introduction}

Dark matter could be mixed with ordinary nuclear matter inside neutron stars in sufficient quantities to affect bulk properties of the star, such as the star's mass and radius.  Such mixed stars have been studied extensively, with dark matter modeled as a bosonic \cite{Henriques:1989ar, Henriques:1989ez, Henriques:1990xg, deSousa:1995ye, Pisano:1995yk, Sakamoto:1998aj, deSousa:2000eq, Henriques:2003yr, Dzhunushaliev:2011ma, ValdezAlvarado:2012xc, Brito:2015yga, Brito:2015yfh, Bezares:2019jcb, DiGiovanni:2020frc, Valdez-Alvarado:2020vqa, DiGiovanni:2021vlu, Kain:2021bwd, Karkevandi:2021ygv, Lee:2021yyn, DiGiovanni:2021ejn, Sanchis-Gual:2022ooi} or fermionic \cite{Sandin:2008db, Ciarcelluti:2010ji, Goldman:2011aa, Leung:2011zz, Leung:2012vea, Li_2012, Li:2012ii, Leung:2013pra, Goldman:2013qla, Xiang:2013xwa, Tolos:2015qra, Mukhopadhyay:2015xhs, Panotopoulos:2017pgv, Panotopoulos:2017idn, Gresham:2018rqo, Nelson:2018xtr, Ellis:2018bkr, Deliyergiyev:2019vti, Bhat:2019tnz, DelPopolo:2020pzh, Zhang:2020dfi, Das:2020vng, Kain:2020zjs, Kain:2021hpk, Das:2021dru, Das:2021yny, Sen:2021wev, Das:2021wku, Das:2020ptd, Jimenez:2021nmr, kain, MafaTakisa:2021fui, Chan:2021gcm, Das:2021hnk, Lourenco:2021dvh, Ryan:2022hku, Miao:2022rqj} particle.  Bosonic dark matter has most commonly been modeled with a complex scalar field.  In this case, the mixed star is called a fermion-boson star, and is the subject of this work.

Much of the literature on mixed stars is devoted to the study of static solutions, for which spacetime is time independent.  Dynamical solutions, on the other hand, are valuable, since they study nonlinear effects and can determine if the system is capable of migrating from one configuration to another.  Dynamical evolutions of mixed stars with bosonic dark matter were presented in \cite{ValdezAlvarado:2012xc, Brito:2015yga, Brito:2015yfh, Bezares:2019jcb, DiGiovanni:2020frc, Valdez-Alvarado:2020vqa, DiGiovanni:2021vlu} and with fermionic dark matter in \cite{kain}.  One shortcoming of these papers is that they described ordinary nuclear matter with a simple polytropic equation of state.  A primary purpose of this work is to present the first dynamical evolution of a mixed star using a realistic equation of state, one that includes all fermions expected to exist in the core, accounts for beta-equilibrium and charge neutrality, and includes inner and outer crusts.

Complex scalar fields, by themselves, can form starlike configurations known as boson stars \cite{Kaup:1968zz, Ruffini:1969qy, Colpi:1986ye, Seidel:1991zh, Liebling:2012fv, Kain:2021rmk}.  In a seminal paper, Seidel and Suen made a detailed study of the dynamical evolution of both stable and unstable static solutions of boson stars subject to weak and strong perturbations \cite{Seidel:1990jh}.  Their framework has since been used in similar studies of related systems, such as oscillatons \cite{Alcubierre:2003sx, UrenaLopez:2012zz}, Proca stars \cite{Sanchis-Gual:2017bhw}, and Dirac stars \cite{Daka:2019iix}.  Another purpose of this work is to present a similar study of fermion-boson stars.

For weak perturbations, our dynamical evolutions show that linearly stable static solutions are also nonlinearly stable.  For unstable static solutions, we identify the regions of parameter space which collapse to black holes and which migrate to the stable region.  For simplicity, we consider only strong perturbations that increase the total mass of the system.  We find that unstable static solutions that migrate to the stable region, when strongly perturbed, collapse to black holes.  For strongly perturbed stable static solutions, we identify the criteria under which the initial configuration moves to the stable region and does not collapse.

In the next section, we present the equations that we solve numerically for dynamical and static solutions, as well as describe our numerical methods.  In Sec.\ \ref{sec:eos}, we discuss the realistic equations of state our code is able to evolve.  In Sec.\ \ref{sec:stability}, we present critical curves, which identify linearly stable static solutions of fermion-boson stars over the whole of parameter space.  Our main results begin in Sec.\ \ref{sec:weak}, where we dynamically evolve weakly perturbed static solutions.  In Sec.\ \ref{sec:strong}, we present results for strongly perturbed static solutions.  We conclude in Sec.\ \ref{sec:conclusion}.


\section{Fermion-boson stars}

In this section, we derive equations describing spherically symmetric fermion-boson stars.  This includes equations for both dynamical and static solutions.  We use units such that $c=\hbar =1$ and parametrize the spherically symmetric metric as
\begin{equation} \label{metric}
ds^2 = -\alpha^2 dt^2 + a^2 dr^2 + r^2 (d\theta^2 + \sin^2\theta d\phi^2),
\end{equation}
where $\alpha$ and $a$ are metric functions determined from the Einstein field equations, $G\indices{^\mu_\nu} = 8\pi G (T_\text{tot})\indices{^\mu_\nu}$, where $G\indices{^\mu_\nu}$ is the Einstein tensor, $G = 1/m_P^2$ where $m_P$ the Planck mass, and $(T_\text{tot})\indices{^\mu_\nu}$ is the energy-momentum tensor.

The energy-momentum tensor contains contributions from both the fermion and boson sectors.  We make the standard assumption that nongravitational interactions between these sectors are negligible, so that the energy-momentum tensor takes the form
\begin{equation}
(T_\text{tot})\indices{^\mu_\nu} = (T_f)\indices{^\mu_\nu} + (T_b)\indices{^\mu_\nu},
\end{equation}
where the fermionic and bosonic contributions are independently conserved,
\begin{equation}
\nabla_\mu (T_f)\indices{^\mu_\nu} = 0, \qquad
\nabla_\mu (T_b)\indices{^\mu_\nu} = 0,
\end{equation}
with $\nabla_\mu$ being the metric-covariant derivative.

The Einstein field equations lead to the following equations for determining $\alpha$ and $a$,
\begin{equation} \label{metric equations}
\begin{split}
\frac{\alpha'}{\alpha} &= +4\pi G r  a^2 (T_\text{tot})\indices{^r_r} + \frac{a^2-1}{2r}
\\
\frac{a'}{a} &= 
-4\pi G r a^2 (T_\text{tot})\indices{^t_t}
- \frac{a^2 - 1}{2r} 
\\
\frac{\dot{a}}{a} &= -4\pi G r \alpha^2 (T_\text{tot})\indices{^t_r},
\end{split}
\end{equation}
where a prime denotes an $r$ derivative and a dot denotes a $t$ derivative.  Since our parametrization of the metric in Eq.\ (\ref{metric}) uses the Schwarzschild radial coordinate, the total mass inside a radius $r$ is given by
\begin{equation} \label{m eq}
m = \frac{r}{2G} \left(1 - \frac{1}{a^2} \right).
\end{equation}


\subsection{Fermion sector}
\label{sec:fermion}

The fermion sector is composed of the ordinary nuclear matter of a neutron star.  We use a hydrodynamical description and describe the fermion sector with a perfect fluid energy-momentum tensor,
\begin{equation} \label{pf em}
(T_f)^{\mu\nu} = (\rho + P)u^\mu u^\nu + P g^{\mu\nu},
\end{equation}
where $u^\mu$ is the four-velocity, $\rho$ is the energy density, and $P$ is the pressure of the fluid, and an equation of state
\begin{equation} \label{eos form}
P = P(\rho).
\end{equation}
We note that we are not allowing for the more general equation of state $P = P(\rho,n_f)$, where $n_f$ is the fluid number density; equations of state of the form (\ref{eos form}) are called barotropic.

In spherically symmetric systems, $u^\theta = u^\phi = 0$.  Defining $v \equiv a u^r/\alpha u^t$, the fermion sector is described by $P$, $\rho$, and $v$, which are known as the \textit{primitive} variables.  To solve for the primitive variables, we write the equations of motion in conservation form.  This requires the introduction of \textit{conservative} variables which, following \cite{Neilsen:1999we, kain}, we take to be
\begin{equation} \label{Pi Phi def}
\Pi \equiv \frac{\rho + P}{1-v} - P, \qquad
\Phi \equiv \frac{\rho + P}{1+v} - P.
\end{equation}
It is impossible to fully invert these formulas and solve for the primitive variables analytically.  Trying to invert them leads to
\begin{align} \label{prim from con}
0 &= (\Pi - \Phi)^2 - (\Pi + \Phi - 2\rho) (\Pi + \Phi + 2P)
\notag \\
v &= \frac{\Pi - \Phi}{\Pi + \Phi + 2P}.
\end{align}
When combined with the equation of state $P(\rho)$, the first equation can be solved numerically for $\rho$, which gives $P$, from which the second equation gives $v$.

The equations of motion follow from conservation of the energy-momentum tensor, $\nabla_\mu (T_f)\indices{^\mu_\nu} = 0$ for $\nu=t,r$, and the metric equations in (\ref{metric equations}).  We write the equations of motion in conservation form,
\begin{equation} \label{f eos}
\partial_t \mathbf{u} 
+\frac{1}{r^2} \partial_r \left(r^2 \frac{\alpha}{a} \mathbf{f}^{(1)} \right) 
+\partial_r \left(\frac{\alpha}{a} \mathbf{f}^{(2)} \right) 
= \mathbf{s},
\end{equation}
where $\mathbf{u}$ is the vector of conservative variables, $\mathbf{f}^{(1)}$ and $\mathbf{f}^{(2)}$ are the vectors of fluxes, and $\mathbf{s}$ is the vector of sources,
\begin{align}
\mathbf{u} &= 
\begin{pmatrix}
\Pi \\ \Phi
\end{pmatrix}&
\mathbf{f}^{(1)} &= 
\begin{pmatrix}
\frac{1}{2}(\Pi - \Phi)( 1 + v) \\
\frac{1}{2}(\Pi - \Phi)( 1 - v)
\end{pmatrix}
\notag \\
\mathbf{f}^{(2)} &= 
\begin{pmatrix}
+ P \\
- P
\end{pmatrix}&
\qquad
\mathbf{s} &= 
\begin{pmatrix}
\Omega + \Theta
\\
\Omega - \Theta
\end{pmatrix},
\end{align}
with the source terms given by
\begin{widetext}
\begin{equation}
\begin{split}
\Omega &= -4\pi G r \alpha^2
\Bigl\{
(T_f)\indices{^t_r}  \left[ (T_\text{tot})\indices{^r_r} - (T_\text{tot})\indices{^t_t} \right]
- (T_\text{tot})\indices{^t_r} [(T_f)\indices{^r_r} - (T_f)\indices{^t_t}]
\Bigr\}
\\
\Theta &= 
\frac{\alpha}{a} \frac{a^2-1}{2r} 
\left\{ \frac{1}{2}
\left[
(\Pi-\Phi)v - (\Pi+\Phi)
\right]+P
\right\}
+4\pi G r a \alpha \Biggl[ 2\frac{\alpha^2}{a^2} (T_\text{tot})\indices{^t_r} (T_f)\indices{^t_r}
+ (T_\text{tot})\indices{^r_r} (T_f)\indices{^t_t}
+  (T_\text{tot})\indices{^t_t} (T_f)\indices{^r_r}
\Biggr].
\end{split}
\end{equation}
\end{widetext}
The relevant components of the energy-momentum tensor, written in terms of a combination of conservative and primitive variables, are
\begin{equation}\label{f em}
\begin{split}
(T_f)\indices{^t_t} &= -\frac{1}{2} (\Pi+\Phi)
\\
(T_f)\indices{^r_r} &= \frac{1}{2}(\Pi-\Phi) v + P
\\
(T_f)\indices{^t_r} &=  \frac{a}{2\alpha} (\Pi-\Phi).
\end{split}
\end{equation}

The fermion sector is a hydrodynamical system.  Dynamical evolution of hydrodynamical systems generate discontinuities, or shocks, that cause the failure of simple finite difference schemes.  To solve the equations of motion in (\ref{f eos}) we use finite volume and high-resolution shock-capturing methods \cite{Romero:1995cn, RezzollaBook}.  In particular, we use the methods developed in \cite{kain, Neilsen:1999we}.  Reference \cite{kain} studied mixed stars with fermionic dark matter by dynamically evolving two-fluid systems.  Here, we use the single-fluid version of these methods \cite{Neilsen:1999we}.  We refer the reader to \cite{kain, Neilsen:1999we} for details.  


\subsection{Boson sector}

The boson sector is composed of bosonic dark matter, which we model with a complex scalar field.  We describe the boson sector with Lagrangian
\begin{equation} \label{L}
\mathcal{L}_b = - (\partial_\mu \phi) (\partial^\mu \phi)^* - V,
\qquad
V = \mu^2 |\phi|^2 + \lambda |\phi|^4,
\end{equation}
where $\phi$ is the complex scalar field, $\mu$ is its mass, and $\lambda$ is its self-coupling constant.  We minimally couple the scalar field to gravity through $\mathcal{L}_b \rightarrow \sqrt{-\det(g_{\mu\nu})} \mathcal{L}_b$.  From this Lagrangian, it is straightforward to compute the equations of motion,
\begin{equation} \label{eom}
\begin{split}
\partial_t \left( \frac{a}{\alpha} \dot{\phi} \right) &= \frac{1}{r^2} \partial_r \left( r^2 \frac{\alpha}{a} \phi' \right)
- \alpha a \left( \mu^2 + 2 \lambda |\phi|^2 \right) \phi
\end{split}
\end{equation}
and the following components of the energy-momentum tensor:
\begin{equation} \label{em comps}
\begin{split}
(T_b)\indices{^t_t} &= - \frac{1}{\alpha^2} |\dot{\phi}|^2 - \frac{1}{a^2} |\phi'|^2 - V
\\
(T_b)\indices{^r_r} &= + \frac{1}{\alpha^2} |\dot{\phi}|^2 + \frac{1}{a^2} |\phi'|^2 - V
\\
(T_b)\indices{^t_r} &= - \frac{1}{\alpha^2} \left( \dot{\phi} \phi^{*\prime} + \dot{\phi}^* \phi' \right).
\end{split}
\end{equation}

To help with solving the equations of motion, we write fields in terms of their real and imaginary parts and write the equations of motion in first order form.  Defining the (complex) auxiliary fields
\begin{equation} \label{X Y def}
X \equiv \phi', \qquad Y \equiv \frac{a}{\alpha} \dot{\phi},
\end{equation}
and then separating fields into real and imaginary parts,
\begin{equation}
\phi = \phi_1 + i \phi_2, \qquad
X = X_1 + i X_2, \qquad
Y = Y_1 + iY_2,
\end{equation}
the equations of motion become 
\begin{equation} \label{b evo1}
\begin{split}
\dot{\phi}_1 &= \frac{\alpha}{a} Y_1 
\\
\dot{X}_1 &=
\partial_r
\left(\frac{\alpha}{a} Y_1\right) 
\\
\dot{Y}_1 &=
\frac{1}{r^2} \partial_r
\left( \frac{r^2 \alpha}{a}X_1\right)
- \alpha a
[\mu^2 + 2 \lambda (\phi_1^2 + \phi_2^2) ] \phi_1
\end{split}
\end{equation}
and
\begin{equation} \label{b evo2}
\begin{split}
\dot{\phi}_2 &= \frac{\alpha}{a} Y_2 
\\
\dot{X}_2 &=
\partial_r
\left(\frac{\alpha}{a} Y_2 \right) 
\\
\dot{Y}_2 &=
\frac{1}{r^2} \partial_r
\left( \frac{r^2 \alpha}{a}X_2 \right)
- \alpha a
[\mu^2 + 2 \lambda (\phi_1^2 + \phi_2^2) ] \phi_2
\end{split}
\end{equation}
and the energy-momentum tensor components in (\ref{em comps}) become
\begin{equation} \label{b em}
\begin{split}
(T_b)\indices{^t_t} 
&= -\frac{X_1^2 + X_2^2 + Y_1^2 + Y_2^2}{a^2}
- V
\\
(T_b)\indices{^r_r} 
&=+\frac{X_1^2 + X_2^2 + Y_1^2 + Y_2^2}{a^2}
- V
\\
(T_b)\indices{^t_r} 
&=
- \frac{2}{\alpha a} \left(X_1 Y_1 + X_2 Y_2 \right),
\end{split}
\end{equation}
where $V = \mu^2(\phi_1^2 + \phi_2^2) + \lambda(\phi_1^2 + \phi_2^2)^2$.

Two parameters in the boson sector are the boson particle mass, $\mu$, and the self-coupling constant, $\lambda$.  For completeness, we present equations for arbitrary $\lambda$, but for simplicity, we present results for $\lambda = 0$. A useful dimensionless quantity for understanding the relative mass scales for the fermion and boson sectors can be shown to be \cite{Henriques:1989ar}
\begin{equation}
\beta \equiv \frac{(\mu_N/m_P)^2}{\mu/m_P},
\end{equation}
where $m_P$ is the Planck mass and $\mu_N$ is the representative particle mass for the fermion sector, which we take to be the nucleon mass $\mu_N = 938$ MeV.  Roughly speaking, when $\beta$ is order one, there is a single mass scale in the system and the fermion and boson sectors can both be non-negligible \cite{Henriques:1989ar}.  For order one $\beta$, we have $\mu \sim 10^{-10}$ eV.  Nearly all studies of fermion-boson stars assume a boson particle mass near this value, and we shall do the same.  In the following, we use $\mu = 10^{-10}$ eV.  For details on how different particle masses affect fermion-boson stars, see, for example, Refs.\ \cite{Henriques:1989ar, Kain:2021bwd, DiGiovanni:2021ejn}.


\subsection{Numerical methods}

We solve the equations of motion in (\ref{f eos}), (\ref{b evo1}), and (\ref{b evo2}) using the method of lines, second order finite differencing in the spatial direction, and third-order Runge-Kutta in the time direction.  In the fermion sector, we use the minmod slope limiter, the Harten-Lax-van Leer-Einfeldt (HLLE) approximate Riemann solver, and we numerically solve the top equation in (\ref{prim from con}) for $\rho$ using the Newton-Raphson method.  We solve for the metric functions using the first two equations in (\ref{metric equations}), second-order Runge-Kutta, the inner boundary condition $a=1$, and the outer boundary condition $\alpha = 1/a$.

Our computational domain does not extend to infinity.  We therefore allow fields to exit at the outer boundary.  In the fermion sector, we use the outer boundary conditions given in \cite{Neilsen:1999we}.  In the boson sector, we use standard outgoing wave boundary conditions,
\begin{equation}
\begin{split}
\dot{\phi}_1 &= -\frac{\phi_1}{r} - X_1, \quad
\dot{Y}_1 = - \frac{Y_1}{r} - Y_1', \quad
X_1 = - \frac{\phi_1}{r} - Y_1,
\\
\dot{\phi}_2 &= -\frac{\phi_2}{r} - X_2, \quad
\dot{Y}_2 = - \frac{Y_2}{r} - Y_2', \quad
X_2 = - \frac{\phi_2}{r} - Y_2.
\end{split}
\end{equation}

Our code uses dimensionless time and space variables defined by
\begin{equation}
\bar{t} \equiv \frac{(1\text{ GeV})^2}{m_P} t, \qquad
\bar{r} \equiv \frac{(1\text{ GeV})^2}{m_P} r.
\end{equation}
Most of the results shown in upcoming sections were computed using uniform grid spacing $\Delta \bar{r} = 0.02$, time step $\Delta \bar{t} = 0.5 \Delta \bar{r}$,  and an outer boundary at $\bar{r}_\text{max} = 350$.  We have occasionally used smaller values for $\Delta\bar{r}$.  We have confirmed that our code is second order convergent.


\subsection{Initial data:\ Static solutions}
\label{sec:static}

In this work, we study the dynamical evolution of both linearly stable and unstable static solutions.  Static solutions are defined as solutions for which spacetime is time independent.  In the fermion sector, this leads to all fields being time independent.  Time independent fields require vanishing fluid velocities and the perfect fluid energy-momentum tensor in (\ref{pf em}) reduces to
\begin{equation}
(T_f)\indices{^\mu_\nu} = \text{diag} (-\rho, P, P, P).
\end{equation}
The equations of motion still follow from $\nabla_\mu (T_f)\indices{^\mu_\nu} = 0$, but now work out to
\begin{equation} \label{P' eq}
P' = - \frac{\alpha'}{\alpha} (\rho + P),
\end{equation}
where $\alpha'/\alpha$ is given by the top equation in (\ref{metric equations}).

In the boson sector, the scalar field must retain a time dependence, even though spacetime is time independent.  Equations for static solutions follow from the boson star ansatz,
\begin{equation} \label{ansatz}
\phi(t,r) = \varphi(r) e^{i\omega t},
\end{equation}  
where $\varphi(r)$ is real and $\omega$ is a real constant called the oscillation frequency.  With this ansatz, the energy-momentum tensor components reduce to
\begin{equation} 
\begin{split}
(T_b)\indices{^t_t}
&= -\frac{\omega^2}{\alpha^2} \varphi^2 - \frac{\varphi^{\prime \, 2}}{a^2}
- 
\mu^2 \varphi^2 - \lambda \varphi^4
\\
(T_b)\indices{^r_r} &= +\frac{\omega^2}{\alpha^2} \varphi^2 + \frac{\varphi^{\prime\,2}}{a^2}
-
\mu^2 \varphi^2 - \lambda \varphi^4
\end{split}
\end{equation}
and the equations of motion reduce to
\begin{equation} \label{bs eom 0}
\varphi'' = \varphi' \left( \frac{a'}{a} - \frac{\alpha'}{\alpha} - \frac{2}{r} \right)
- \frac{a^2 \omega^2}{\alpha^2}\varphi
+ a^2 \left( \mu^2 \varphi + 2\lambda \varphi^3 \right),
\end{equation}
where $a'/a$ is given by the middle equation in (\ref{metric equations}).

In the absence of dark matter, a static solution for a fermion star is found by solving Eq.\ (\ref{P' eq}) and the top two equations in (\ref{metric equations}), which together are known as the Tolman-Oppenheimer-Volkoff (TOV) equations.  When the top equation in (\ref{metric equations}) is plugged into Eq.\ (\ref{P' eq}), $\alpha$ decouples and does not have to be solved for.   Each static solution is uniquely identified by the central pressure, $P(0)$.  Thus, after specifying a central pressure, Eq.\ (\ref{P' eq}) and the middle equation in (\ref{metric equations}) can be integrated outward from $r=0$.  The edge of the star, at $r=R$, is defined by $P(R) = 0$ and the mass of the star is given by $M = m(R)$, where $m$ is given in Eq.\ (\ref{m eq}).

In the presence of dark matter, static solutions for fermion-boson stars are found by solving Eqs.\ (\ref{P' eq}) and (\ref{bs eom 0}) along with the top two equations in (\ref{metric equations}).  Static solutions are uniquely identified by the central pressure in the fermion sector, $P(0)$, and the central scalar field value in the boson sector, $\varphi(0)$.  After specifying these values, as well as a trial value for the oscillation frequency $\omega$, the equations can be integrated outward from $r= 0$.  We then vary $\omega$ using the shooting method until the outer boundary condition $\varphi \rightarrow 0$ as $r \rightarrow \infty$ is satisfied.  The total mass of the system, $M$, which includes contributions from both sectors, is given by $M = m(\infty)$.


\section{Equations of state}
\label{sec:eos}

Our description of the fermion sector makes use of an equation of state, $P(\rho)$, which describes the ordinary nuclear matter of a neutron star.  Since the precise properties of nuclear matter at the extreme pressures found in the core of a neutron star are unknown, the correct equation of state is also unknown.  Consequently, many equations of state have been proposed.  

It is valuable to use a realistic equation of state, which includes all particles expected to exist in the core, accounts for beta-equilibrium and charge neutrality, and includes an inner and outer crust.  Studies of the static solutions of mixed stars with fermionic dark matter have commonly adopted realistic equations of state for ordinary matter (see, for example, Refs.\ \cite{Sandin:2008db, Leung:2011zz, Tolos:2015qra, Kain:2021hpk}).  Studies of the static solutions of mixed stars with bosonic dark matter have only recently done so \cite{Kain:2021bwd, Karkevandi:2021ygv, Lee:2021yyn, DiGiovanni:2021ejn}.

Dynamical evolutions of mixed stars with bosonic dark matter \cite{ValdezAlvarado:2012xc, Brito:2015yga, Brito:2015yfh, Bezares:2019jcb, DiGiovanni:2020frc, Valdez-Alvarado:2020vqa, DiGiovanni:2021vlu} and fermionic dark matter \cite{kain} have so far exclusively used a simple polytropic equation of state to describe ordinary matter.  A primary purpose of this work is to present for the first time dynamical evolutions of mixed stars using a realistic equation of state.  

Using a realistic equation of state is complicated by the fact that dynamical evolutions require smooth equations to retain numerical stability, but many equations of state are presented in a tabulated form.  We overcome this by using analytical fits \cite{Haensel:2004nu, Potekhin:2013qqa} of tabulated equations of state.  In particular, our code is able to evolve the analytical fit \cite{Haensel:2004nu} of the FPS \cite{FPS} and SLy \cite{Douchin:2001sv} equations of state and the analytic fit \cite{Potekhin:2013qqa} of the BSk19, BSk20, and BSk21 \cite{Goriely:2010bm, Pearson:2011zz, Pearson:2012hz} equations of state.  All five of these are unified equations of state, in that they were originally calculated with a single effective nuclear Hamiltonian, allowing for a smooth transition between crusts and between the inner crust and core.  This additional level of smoothness, on top of that obtained via the analytical fit, helps retain numerical stability in our dynamical evolutions.

\begin{figure}
\centering
\includegraphics[width=3.4in]{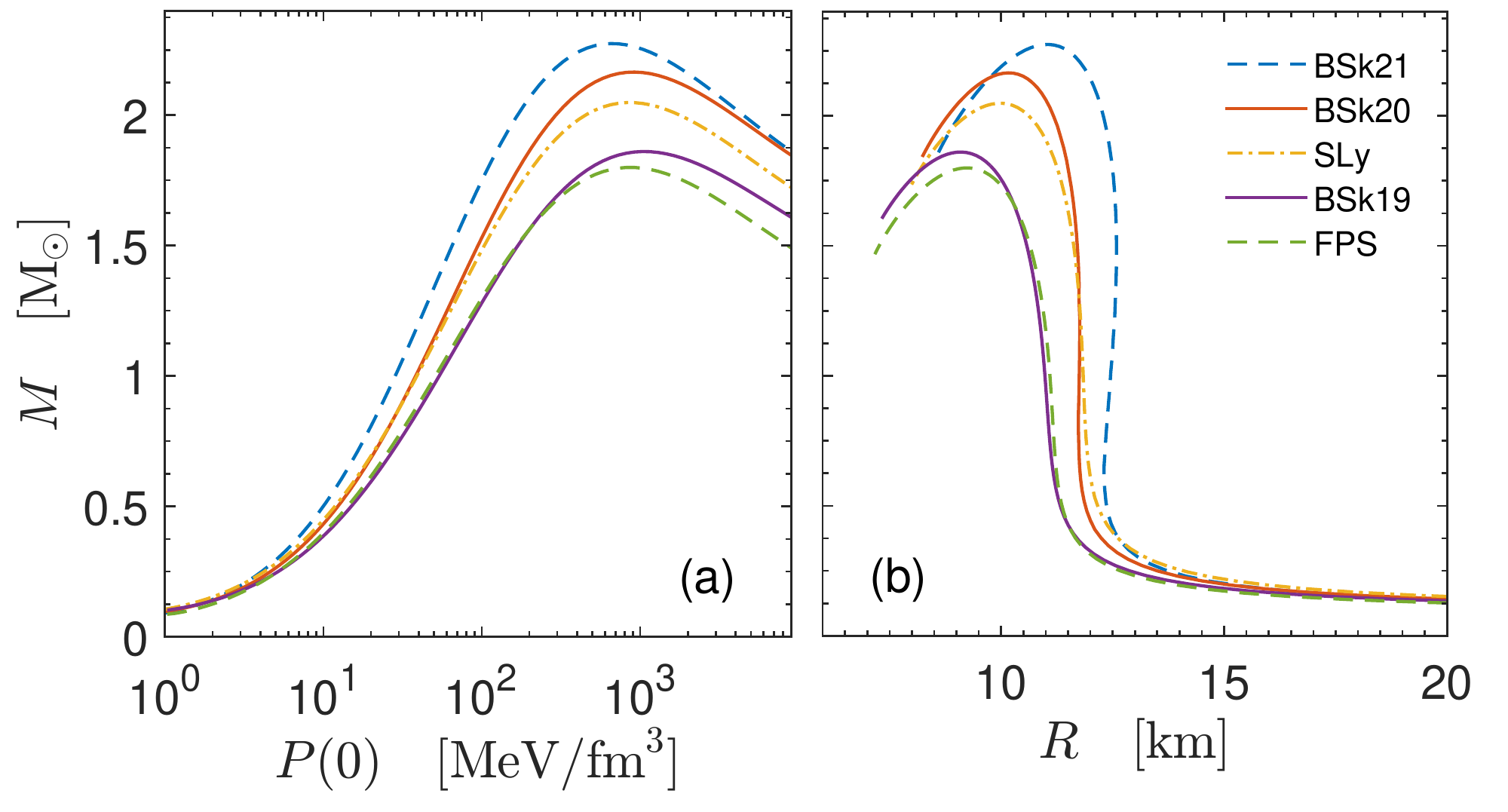}
\caption{Each point on a curve represents a static solution for a neutron star, as computed from the TOV equations using one of the equations of state listed in the legend.  The two figures plot the same solutions, with (a) displaying the mass of the star, $M$, as a function of the central pressure, $P(0)$, and (b) displaying the mass as a function of the radius of the star, $R$.}
\label{fig:eos}
\end{figure}

We have solved the TOV equations for static solutions using each of the equations of state.  These static solutions describe fermion stars in the absence of dark matter, i.e.\ neutron stars.  The results are shown in Fig.\ \ref{fig:eos}.  Each point on a curve in Fig.\ \ref{fig:eos} represents a static solution.  It is well-known that the static solution with the largest mass, which is called the critical solution, marks the transition from linearly stable to unstable:\ static solutions with central pressures smaller than the critical pressure are linearly stable; otherwise they are unstable.  We list properties of the critical solutions in Table \ref{1fluid critical}.

\begin{table}
\begin{tabular}{c@{\hskip 5pt}|@{\hskip 5pt}c@{\hskip 5pt}|@{\hskip 5pt}c@{\hskip 5pt}|@{\hskip 5pt}c}
eos & $M$ [M$_\odot$] & R [km] & $P(0)$ [MeV/fm$^3$] \\
\hline
FPS & 1.800 & 9.234 & 866.0 \\
BSk19 & 1.861 & 9.090 & 1061.5 \\
SLy & 2.048 & 9.979 & 860.3 \\
BSk20 & 2.165 & 10.175 & 906.1 \\
BSk21 & 2.274 & 11.038 & 665.3
\end{tabular}
\caption{The critical values for neutron stars as computed from the TOV equations using one of the listed equations of state (eos).  Critical values are the values of parameters for the critical solution, which is the static solution with the largest mass (cf.\ Fig.\ \ref{fig:eos}).  The critical solution marks the transition between linearly stable and unstable static solutions.}
\label{1fluid critical}
\end{table}

Spherically symmetric systems made of just a complex scalar field can form static configurations known as boson stars \cite{Liebling:2012fv, Kain:2021rmk}.  The transition from linearly stable to unstable is marked by the static solution with the largest mass, just as it is for fermion stars.  For completeness, we give the critical values for boson stars (with $\mu = 10^{-10}$ eV and $\lambda = 0$) \cite{Kain:2021rmk}:
\begin{equation}
\varphi(0)/m_P = 0.0541, \qquad
M = 0.846 \text{ M}_\odot.
\end{equation}


\section{Stability}
\label{sec:stability}

In Sec.\ \ref{sec:static}, we explained how we solve for static solutions of fermion-boson stars.  Once a solution is found, an immediate question is whether the solution is stable.  In this section, we determine the linear stability of static solutions. 

The determination of linear stability in single-fluid systems is straightforward, since the transition from linearly stable to unstable occurs at the static solution with the largest mass.  Since single-fluid static solutions are uniquely identified by a single parameter, the static solution with the largest mass occupies a single point in parameter space, which is called the critical point.

Static solutions for fermion-boson stars are identified by two parameters, which are the central pressure, $P(0)$, in the fermion sector, and the central scalar field value, $\varphi(0)$, in the boson sector.  The transition from linearly stable to unstable is then marked by a curve in parameter space, which is called the critical curve.  One method for computing the critical curve is to write each field as a harmonic perturbation about its static value and then to solve the linearized system of equations.  This has been done for fermion stars \cite{Chandrasekhar:1964zz}, boson stars \cite{Gleiser:1988rq, Jetzer:1988vr, Gleiser:1988ih, Lee:1988av, Hawley:2000dt, Kain:2021rmk}, mixed stars with fermionic dark matter \cite{Comer:1999rs, Kain:2020zjs}, but not mixed stars with bosonic dark matter, i.e.\ fermion-boson stars.

\begin{figure*}
\centering
\includegraphics[width=7in]{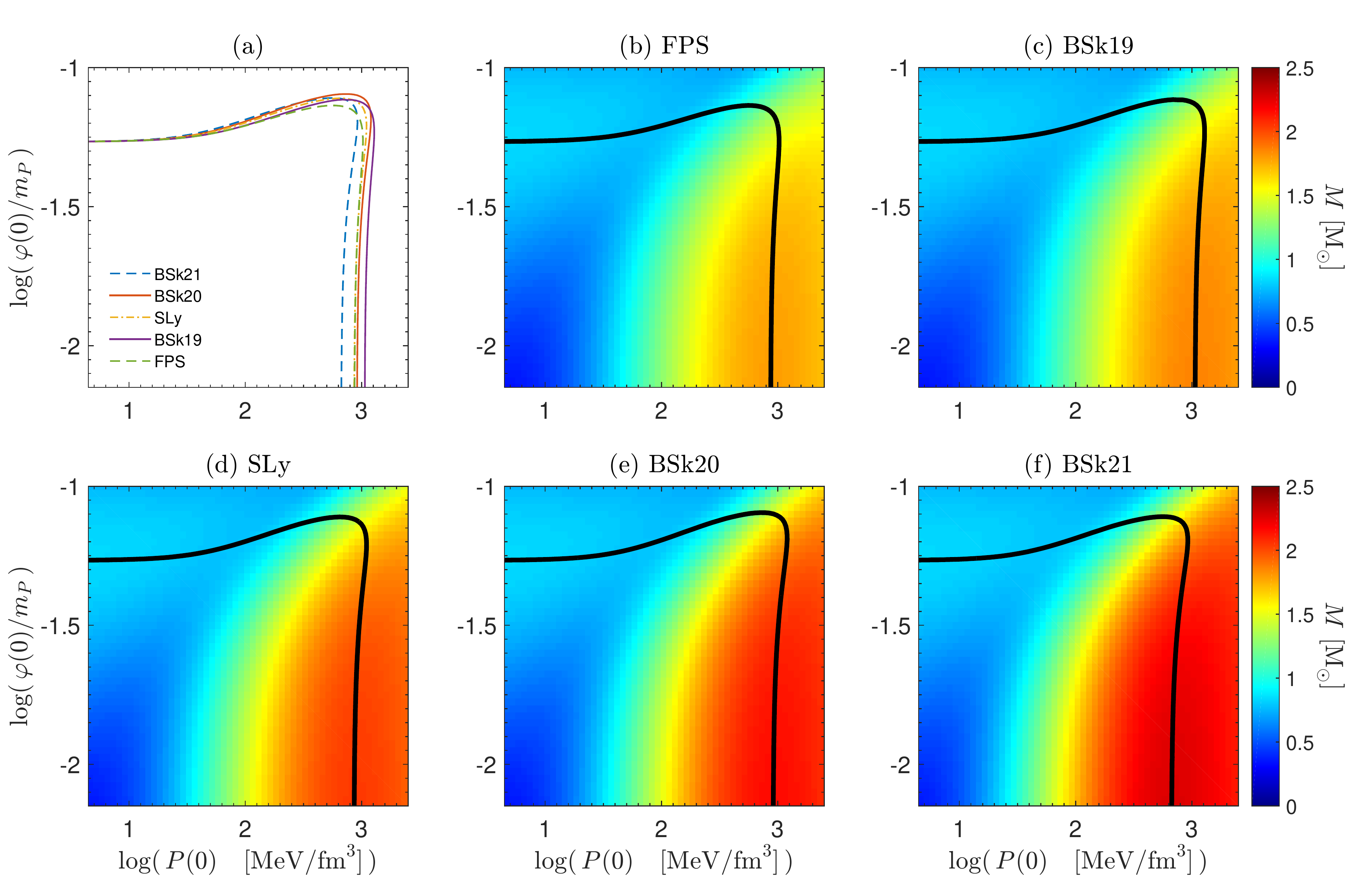}
\caption{Each point in a plot represents a static solution describing a fermion-boson star, which is uniquely identified by the central pressure, $P(0)$, in the fermion sector, and the central scalar field value, $\varphi(0)$, in the boson sector.  The legend in (a) and the titles above (b)--(f) indicate the equation of state used in the fermion sector.   The five curves in (a) and the thick black curves in (b)--(f) are critical curves.  Static solutions below and to the left of their respective critical curve are linearly stable; otherwise they are unstable.  (b)--(f) are also density plots, showing the total mass of the static solution.}
\label{fig:stability}
\end{figure*}

A simpler method for computing critical curves was presented in \cite{Henriques:1990xg}.  Critical curves mark the transition from linearly stable to unstable with respect to perturbations that conserve mass and particle number, and therefore satisfy
\begin{equation} \label{critical curve}
\frac{d M}{d\mathbf{p}} = \frac{d N_f}{d\mathbf{p}} = \frac{d N_b}{d\mathbf{p}} = 0,
\end{equation}
where $M$ is the mass of the system, $N_f$ and $N_b$ are the conserved fermionic and bosonic particle numbers, and $\mathbf{p}$ is a vector in parameter space.  For details on the derivation of this result, we refer the reader to \cite{Henriques:1990xg, ValdezAlvarado:2012xc, Kain:2021hpk}.  It can be shown that if two of the quantities in Eq.\ (\ref{critical curve}) are zero, then the third is also \cite{Henriques:1990xg, Jetzer:1990xa}.  

Before explaining how we solve Eq.\ (\ref{critical curve}), we review how $N_f$ and $N_b$ are computed.  We stress that $N_f$ is the number of fluid elements and not, say, the number of baryons.  From the thermodynamic identity at zero temperature, $\rho + P - \mu n_f = 0$, where $\mu = d\rho/d n_f$ is the chemical potential, the fluid number density, $n_f$, can be computed directly from the equation of state,
\begin{equation} \label{nf eos}
n_f \propto \exp\left[\int \frac{d \rho}{\rho + P(\rho)} \right], 
\end{equation}
where the proportionality constant does not affect the determination of stability and can therefore be arbitrarily chosen.  The total number of fermionic fluid elements inside a radius $r$, $\mathcal{N}_f(r)$, obeys
\begin{equation} \label{Nf eq}
\mathcal{N}_f' = 4\pi a r^2 n_f,
\end{equation}
and $N_f = \mathcal{N}_f(\infty)$.  

In the boson sector, the Lagrangian in (\ref{L}) is invariant under a global phase transformation.  This leads to a conserved current,
\begin{equation} \label{conserved current}
J^\mu = i g^{\mu\nu} (\phi^*\partial_\nu \phi - \phi \partial_\nu \phi^*),
\end{equation}
which satisfies a continuity equation, $\nabla_\mu J^\mu = 0$.  This continuity equation immediately leads to a conserved charge, which is $N_b$.  The total bosonic particle number inside a radius $r$, $\mathcal{N}_b(r)$, is typically written in terms of an integral, $\mathcal{N}_b = \int_0^r d^3r \sqrt{-\det (g_{\mu\nu})} J^t$.  For static solutions, we use the boson star ansatz in (\ref{ansatz}) and prefer to write the integral formula in differential form,
\begin{equation} \label{Nb static}
\mathcal{N}_b' = 8\pi \omega\frac{a \textbf{}}{\alpha} r^2 \varphi^2,
\end{equation}
from which $N_b = \mathcal{N}_b (\infty)$.

To solve Eq.\ (\ref{critical curve}) for the critical curve, we follow the methods presented in \cite{ValdezAlvarado:2012xc, Kain:2021hpk}.  We compute contour lines of either $N_f$ or $N_b$ in the full system.  Moving along a single contour line, we determine the point where $M$ is a maximum.  These points give the critical curve.  

In Fig.\ \ref{fig:stability}, we show the critical curves for the five equations of state we are considering.  Each point in a plot in Fig.\ \ref{fig:stability} represents a static solution.  Those static solutions enclosed by the critical curve (i.e.\ below and to the left of the curve) are linearly stable; otherwise they are unstable.  Figure \ref{fig:stability}(a) shows all five critical curves in one plot.  Figures \ref{fig:stability}(b)--(f) are density plots showing the mass for each static solution, along with the critical curve as the thick black line.  For sufficiently small $P(0)$ or $\varphi(0)$, the system is, respectively, boson or fermion dominated.  In these cases, the critical curves reproduce the single-fluid critical values given in Sec.\ \ref{sec:eos}, as expected.  Interestingly, in the upper-right corner we see that linearly stable static solutions exist for values of $P(0)$ and $\varphi(0)$ that go beyond their single-fluid critical values, a phenomenon that also occurs for mixed stars with fermionic dark matter \cite{Kain:2021hpk}.


\section{Weak perturbations}
\label{sec:weak}

In this section, we dynamically evolve weakly perturbed fermion-boson stars.  Our interest is in determining if linearly stable static solutions are also nonlinearly stable with respect to small perturbations and in determining how unstable static solutions will evolve in time.  To do this, we use static solutions as initial data.  As is well-known, discretization error acts as a weak or small perturbation \cite{Seidel:1990jh}.  For concreteness, starting with this section we focus on the SLy equation of state.  We do not choose SLy for a particular reason and we expect qualitatively similar results with the other equations of state.

\begin{figure}
\centering
\includegraphics[width=3.4in]{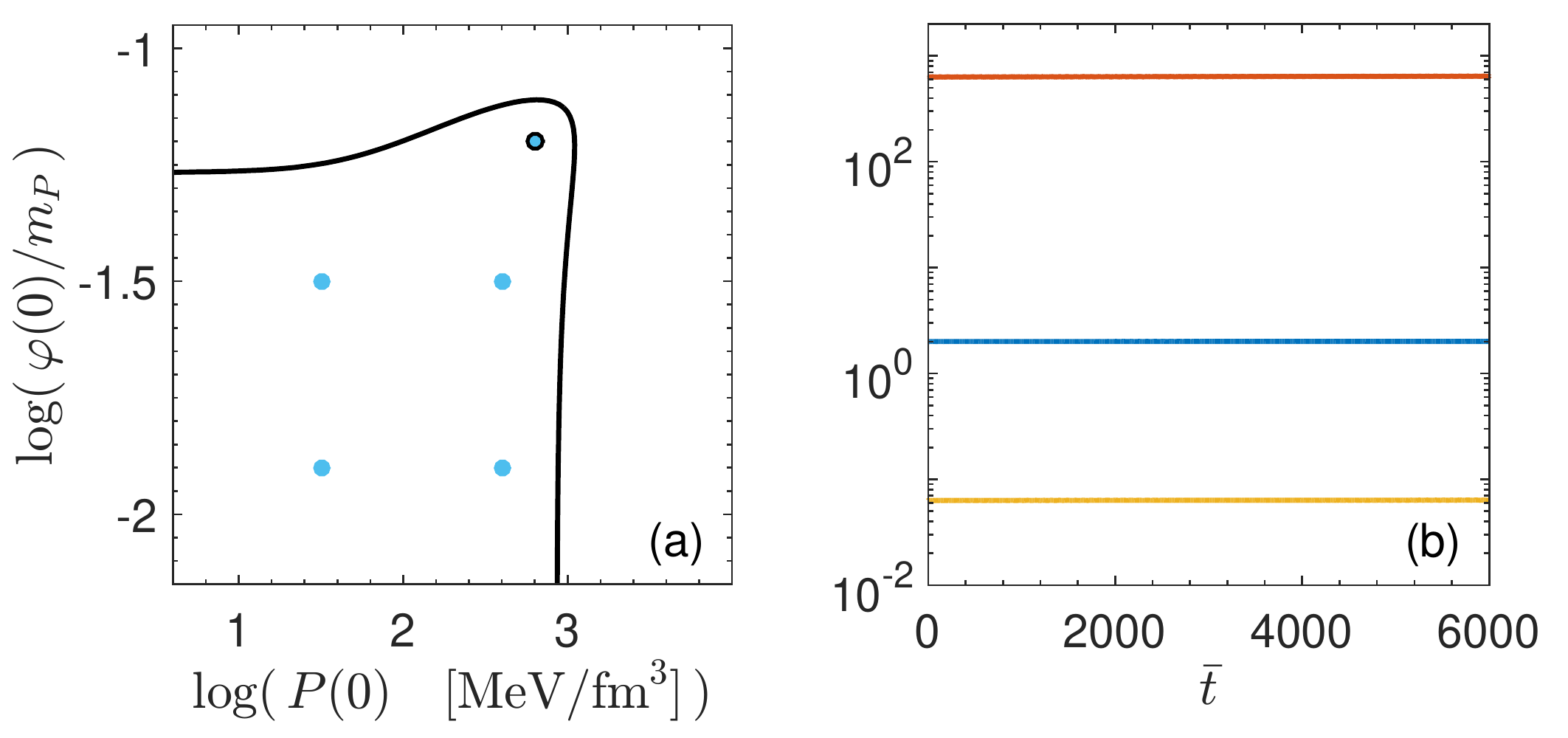}
\caption{(a) The thick black curve is the same critical curve for the SLy equation of state shown in Fig.\ \ref{fig:stability}(d).  The five blue dots indicate representative linearly stable static solutions whose properties are given in Table \ref{table:dynamic stable}.  (b) The dynamical evolution of the weakly perturbed static solution indicated by the blue dot with a black outline in (a).  From top to bottom, the curves display the central pressure, $P(\bar{t},0)$ [MeV/fm$^3$], in the fermion sector (red), $a^2_\text{max}(\bar{t})$ (blue), which is the maximum value of metric component $g_{rr} = a^2$, and the central value of the scalar field, $\varphi(\bar{t},0)/m_P$, in the boson sector (yellow).  That the curves remain flat indicates that the linearly stable static solution is also nonlinearly stable.}
\label{fig:dynamic stable}
\end{figure}

In Fig.\ \ref{fig:dynamic stable}(a), we indicate five representative linearly stable static solutions with blue dots and give their properties in Table \ref{table:dynamic stable}.  We have dynamically evolved each of these static solutions and found each to be nonlinearly stable.  Figure \ref{fig:dynamic stable}(b) shows the dynamical evolution of the static solution indicated by the blue dot with a black outline in Fig.\ \ref{fig:dynamic stable}(a).  That the curves remain flat tells us that the static solution is nonlinearly stable.  We find analogous results for the other static solutions indicated in Fig.\ \ref{fig:dynamic stable}(a).

\begin{table}
\begin{tabular}{c@{\hskip 5pt}|@{\hskip 5pt}c@{\hskip 5pt}|@{\hskip 5pt}c@{\hskip 5pt}|@{\hskip 5pt}c}
$P(0)$ [MeV/fm$^3$] & $\varphi(0)/m_P$ & $M$ [M$_\odot$] & $a^2_\text{max}$ \\
\hline
$10^{1.5}$ & $10^{-1.9}$ & 0.834 & 1.320  \\
$10^{1.5}$ & $10^{-1.5}$ & 0.733 & 1.284  \\
$10^{2.6}$ & $10^{-1.9}$ & 1.946 & 2.280 \\
$10^{2.6}$ & $10^{-1.5}$ & 1.776 & 2.134 \\
$10^{2.8}$ & $10^{-1.2}$ & 1.481 & 2.004 
\end{tabular}
\caption{Properties of the five static solutions indicated by blue dots in Fig.\ \ref{fig:dynamic stable}.}
\label{table:dynamic stable}
\end{table}

\begin{figure*}
\centering
\includegraphics[width=7in]{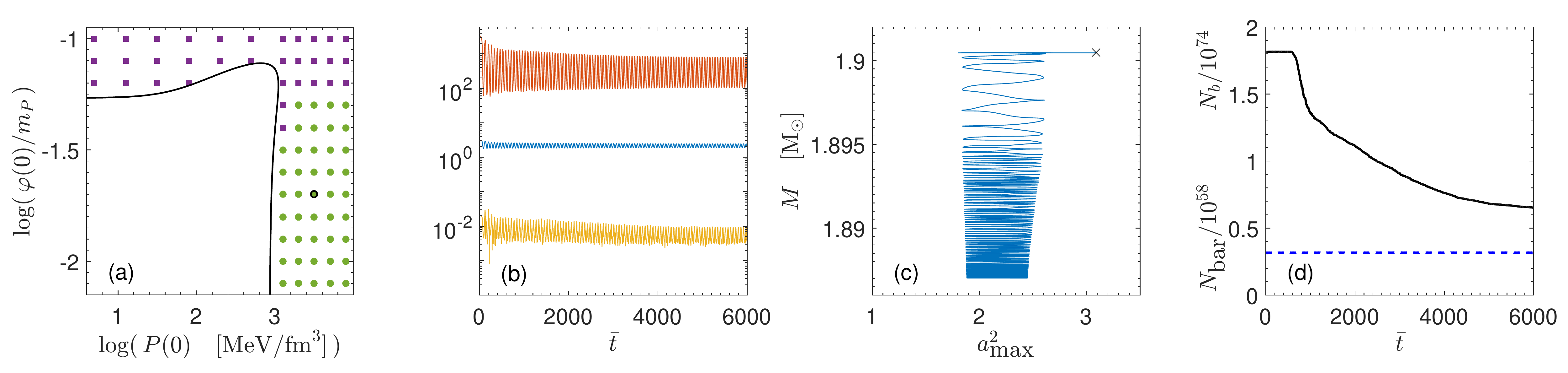}
\caption{(a) The thick black curve is the same critical curve for the SLy equation of state shown in Fig.\ \ref{fig:stability}(d).  The purple squares and green circles indicate weakly perturbed unstable static solutions we evolved.  A purple square indicates that the system collapses to a black hole, while a green circle indicates that the system migrates to a stable configuration.  The dynamical evolution of the green circle with a black outline is shown in (b)--(d).  (b) From top to bottom, the curves display the central pressure, $P(\bar{t},0)$ [MeV/fm$^3$], in the fermion sector (red), $a^2_\text{max}(\bar{t})$ (blue), which is the maximum value of metric component $g_{rr} = a^2$, and the central value of the scalar field, $\varphi(\bar{t},0)/m_P$, in the boson sector (yellow).  (c) The total mass inside the computational domain, $M$, as a function of $a^2_\text{max}$.  (d) The total number of baryons (dashed blue) and bosons (solid black) in the computational domain as a function of time. }
\label{fig:dynamic unstable}
\end{figure*}

Weakly perturbed unstable static solutions will evolve to some other configuration.  We expect three possible outcomes:\ the system migrates to a stable configuration, the system collapses to a black hole, or the system dissipates all matter to infinity.  We have not found evidence for the third possibility, but have found that the system can both migrate and collapse.  In Fig.\ \ref{fig:dynamic unstable}(a), we show the unstable static solutions we have evolved.  The purple squares indicate static solutions that collapse to a black hole.  We determine that a black hole has formed by a spike in the metric function $a$ and the collapse of the metric function $\alpha$ about $r= 0$.  The green circles indicate static solutions that migrate to stable configurations.

To gain insight into how some solutions migrate, we show in Figs.\ \ref{fig:dynamic unstable}(b)--(d) the dynamical evolution of a static solution defined by $P(0) = 10^{3.5}$ MeV/fm$^3$ and $\varphi(0)/m_P = 10^{-1.7}$, which corresponds to the green circle with a black outline in Fig.\ \ref{fig:dynamic unstable}(a).  Figure \ref{fig:dynamic unstable}(b) shows the evolution of a number of fields as a function of time.  As time moves forward, the fields undergo relatively large oscillations.  The top red curve plots the central pressure of the fermion sector, $P(\bar{t}, 0)$.  The time average, over the whole evolution, of the red curve is $10^{2.65}$ MeV/fm$^3$, which is in the stable region, as can be seen from Fig.\ \ref{fig:dynamic unstable}(a).  The bottom yellow curve plots the central scalar field value, $\varphi(\bar{t}, 0)/m_P$.  The time average, over the whole evolution, of the yellow curve is $10^{-2.18}$, which is also in the stable region.

In similar studies of related systems \cite{Seidel:1990jh, Alcubierre:2003sx, UrenaLopez:2012zz, Sanchis-Gual:2017bhw, Daka:2019iix}, tracking the maximum value of the metric component $g_{rr} = a^2$ proved useful.  $a^2_\text{max}$ acts as a probe of the system since it directly depends on all matter.  If a black hole forms, we would see $a^2_\text{max} \rightarrow \infty$, while if all matter dissipates and spacetime becomes flat, we would see $a^2_\text{max} \rightarrow 1$.  We note also that tracking $a^2_\text{max}$ is straightforward to do in a dynamical evolution.  $a^2_\text{max}$ is plotted as a function of time as the middle blue curve in Fig.\ \ref{fig:dynamic unstable}(b), where we can see it moving to neither extreme.  In Fig.\ \ref{fig:dynamic unstable}(c), we show the total mass inside the computational domain, $M$, as a function of $a^2_\text{max}$ for the same evolution shown in Fig.\ \ref{fig:dynamic unstable}(b).  The evolution begins at the top of the plot, at the point indicated with a $\times$.  Moving along the curve is equivalent to moving forward in time.  As time moves forward, the total mass decreases.  At the bottom of the plot, which corresponds to the evolution at late times, we see the density of the plotted curve increasing.  This shows that mass is decreasing at a slower rate and suggests that the system is equilibrating to a configuration with a static total mass.  Figure \ref{fig:dynamic unstable}(c) looks very similar to figures found in related studies \cite{Alcubierre:2003sx, UrenaLopez:2012zz, Daka:2019iix} and, just as in those studies, we conclude that our system is migrating to the stable region.  We find similar results when dynamically evolving the other unstable static solutions indicated by green circles in Fig.\ \ref{fig:dynamic unstable}(a).

Another useful quantity to compute is $\mathcal{N}_b(t,r)$, which gives the total number of bosons inside a radius $r$ at time $t$.  By writing the time component of the conserved current in Eq.\ (\ref{conserved current}) in terms of the auxiliary field $Y = Y_1 + iY_2$, the equation for $\mathcal{N}_b$ can be written as 
\begin{equation} \label{Nb eq}
\mathcal{N}_b'= 8\pi r^2  (\phi_1 Y_2 - \phi_2 Y_1).
\end{equation}
This form is useful for a dynamical evolution, while the previous form in Eq.\ (\ref{Nb static}) is useful for static solutions.  In Fig.\ \ref{fig:dynamic unstable}(d), we show the total number of bosons inside the computational domain, $N_b(t) = \mathcal{N}_b(t,r_\text{max})$, where $r_\text{max}$ is the edge of the computational domain, as the solid black curve for the same evolution shown in Figs.\ \ref{fig:dynamic unstable}(b) and \ref{fig:dynamic unstable}(c).  It is clear that a significant number of bosons are dissipating to infinity.  In the fermion sector, an analogous quantity is $\mathcal{N}_\text{bar}(t,r)$, which gives the total number of baryons inside a radius $r$ at time $t$.  This quantity may be computed from 
\begin{equation} \label{}
\mathcal{N}_\text{bar}' = 4\pi a r^2 n_\text{bar},
\end{equation}
which follows from Eq.\ (\ref{Nf eq}), where $n_\text{bar}$ is the baryonic number density.  We are able to compute this quantity during a dynamical evolution because analytical fits for $n_\text{bar}$ are supplied in Ref.\ \cite{Haensel:2004nu} (we note that analytical fits of $n_f$, the number density of fluid elements, are not available and that computing $\mathcal{N}_f$ in a dynamical evolution is beyond the scope of our code).   In Fig.\ \ref{fig:dynamic unstable}(d), we show the total number of baryons inside the computational domain, $N_\text{bar}(t) = \mathcal{N}_\text{bar}(t,r_\text{max})$, as the dashed blue curve for the same evolution shown in Figs.\ \ref{fig:dynamic unstable}(b) and \ref{fig:dynamic unstable}(c).  It is clear that baryons do not dissipate to infinity.

To help us gain additional insight, we quickly review the analogous situation with pure boson stars \cite{Seidel:1990jh}.  With boson stars, \textit{all} weakly perturbed unstable static solutions, when dynamically evolved, are expected to migrate to the stable branch.  During migration, the system dissipates bosons to infinity through a process known as gravitational cooling \cite{Seidel:1993zk}.  However, if the initial data include an explicit perturbation that increases the total mass by even a small amount, the system collapses to a black hole.  This may be contrasted with pure fermion stars.  Fermion stars do not exhibit gravitational cooling and the dynamical evolution of an unstable static solution generally leads to collapse and not to migration.

Returning to fermion-boson stars, in the upper-left portion of Fig.\ \ref{fig:dynamic unstable}(a), the system is dominated by the boson sector and we might expect the system to act similarly to a pure boson star.  However, we find the system collapses to a black hole instead of migrating to the stable region.  This can be understood as the fermion sector acting as an explicit perturbation which causes collapse.  In the lower-right portion of Fig.\ \ref{fig:dynamic unstable}(a), the system is dominated by the fermion sector and we might expect the system to act similarly to a pure fermion star.  However, we find that the system migrates to the stable region instead of collapsing to a black hole.  Further, we find significant dissipation of bosons during migration.  This can be understand as gravitational cooling in the boson sector allowing for migration of the system.


\section{Strong perturbations}
\label{sec:strong}

In this section, we strongly perturb static solutions by including an explicit perturbation in the initial data.  There are many possibilities for the perturbation, since it can be applied to different fields and take different forms.  To keep things simple, we use the following procedure.  We first import a static solution for the initial data.  We then change the energy density in the fermion sector according to
\begin{equation}
\rho(0,r) \rightarrow \rho(0,r) + \delta \rho(r),
\end{equation}
where
\begin{equation} \label{perturbation}
\delta \rho(r) = A r^2 e^{-(r-b)^2/c^2}
\end{equation}
is a Gaussian perturbation with constant parameters $A$, $b$, $c$.  We then update the pressure $P(0,r)$ using the equation of state and then update the conservative variables $\Pi(0,r)$ and $\Phi(0,r)$ using their definitions in (\ref{Pi Phi def}).  We keep the initial velocity equal to zero, $v(0,r) = 0$, and make no change to the initial data in the boson sector.  As in related studies \cite{Seidel:1990jh, Alcubierre:2003sx, UrenaLopez:2012zz, Sanchis-Gual:2017bhw, Daka:2019iix}, our focus is not so much on the values of the parameters, but on the effect the perturbation has on the total mass of the system.  In the following, for simplicity, we only consider perturbations that increase the total mass. 

Although we do not modify the initial data in the boson sector, the boson sector is still being perturbed.  This follows because we do not change the auxiliary field $Y$, yet $Y$ depends on the metric fields $a$ and $\alpha$, as can be see in Eq.\ (\ref{X Y def}), which do change.  In effect, then, $\dot{\phi}$ at $t = 0$ is being perturbed.  Further insight can be gained by considering Eq.\ (\ref{Nb eq}) for $\mathcal N_b$, which gives the total number of bosons inside a radius $r$.  Note that Eq.\ (\ref{Nb eq}) does not explicitly depend on $a$ and $\alpha$.  The perturbation can therefore be thought of as a perturbation to the total mass of the system in such a way that the total number of bosons is unchanged.

Before presenting our results, we review strongly perturbed pure boson stars \cite{Seidel:1990jh}.  We already mentioned that for a boson star, a weakly perturbed unstable static solution migrates to the stable branch and a strongly perturbed unstable static solution collapses to a black hole.  The latter situation occurs even for perturbations that increase the mass of the system by as little as one percent.  When strongly perturbed stable static solutions are evolved, there are different possible outcomes.  If the perturbation does not increase the mass of the system beyond the critical mass, then the system migrates to the stable branch \cite{Alcubierre:2003sx}.  If the perturbation does increase the mass beyond the critical mass, then the system may either migrate to the stable branch or collapse.  In this latter case, whether the system migrates or collapses depends on the specifics of the perturbation.

We now present our results.  We begin first with the unstable static solutions indicated by the green circles in Fig.\ \ref{fig:dynamic unstable}(a).  In the previous section, we found that this region, when weakly perturbed, migrates to the stable region.  When strongly perturbed, we find that the system always collapses to a black hole.  This occurs even for perturbations that increase the mass of the system by as little as one percent.

\begin{figure}
\centering
\includegraphics[width=3.4in]{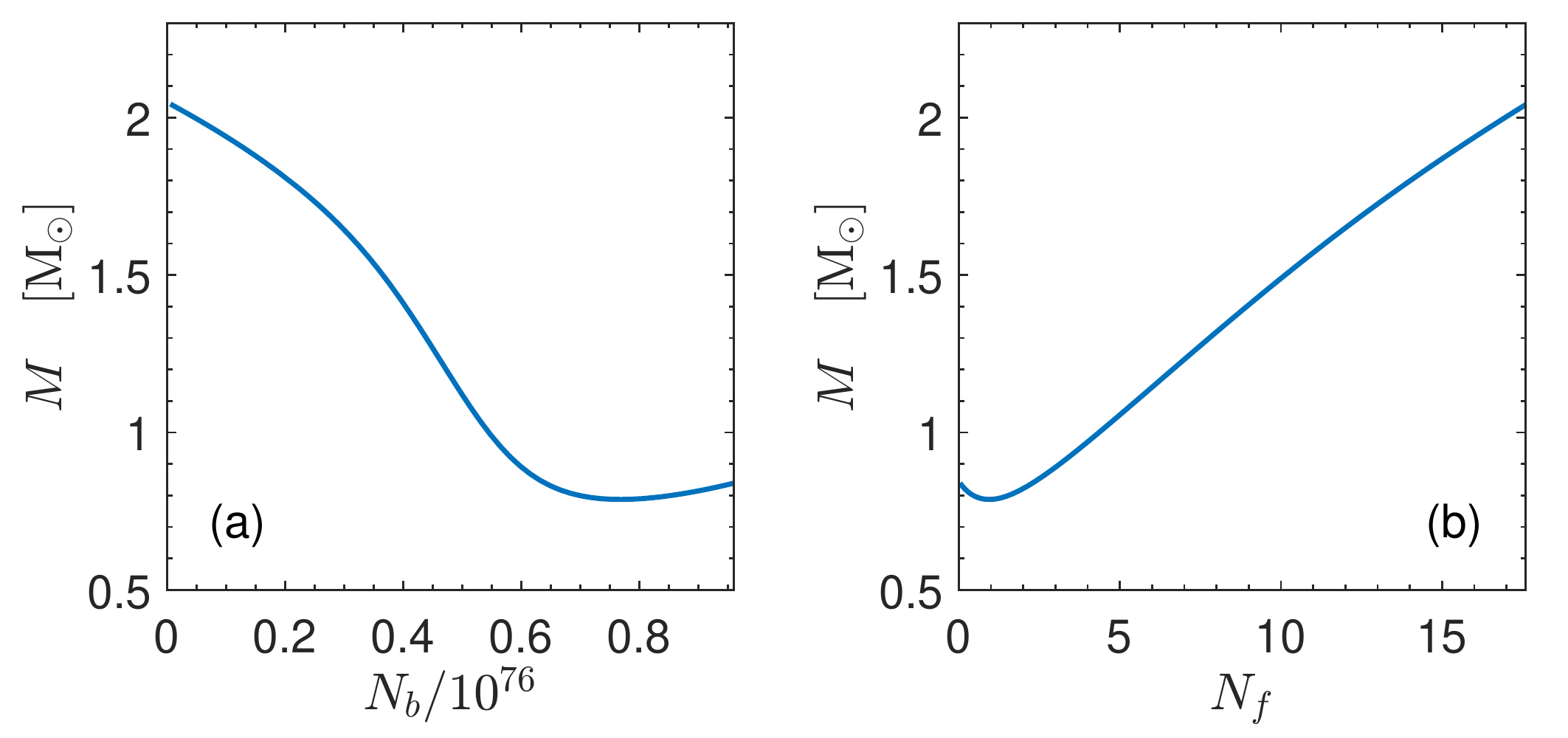}
\caption{These two curves are equivalent to one another and are equivalent to the critical curve shown in Fig.\ \ref{fig:stability}(d).  These curves give the critical mass for a fermion-boson star with the SLy equation of state for a given value of (a) $N_b$, the total number of bosonic particles, and (b) $N_f$, the total number of fluid elements.  Note that the curve in (b) depends upon the arbitrarily chosen proportionality constant in Eq.\ (\ref{nf eos}).}
\label{fig:M Nb}
\end{figure}

The situation is more interesting for strongly perturbed stable static solutions.  We have found the criteria for when such solutions migrate to the stable region and do not collapse.  To understand this criteria, first note that the critical curve is shown as the thick black line in Fig.\ \ref{fig:stability}(d).  In Sec.\ \ref{sec:stability}, we explained the procedure for computing the critical curve:\ Static solutions are defined by the central pressure in the fermion sector, $P(0)$, and the central scalar field value in the boson sector, $\varphi(0)$.  We compute contour lines of either $N_f$, the total number of fluid elements, or $N_b$, the total number of bosons, in the $P(0),\varphi(0)$ parameter space.  Moving along an individual contour line, we find the point where the total mass, $M$, is a maximum.  These points give the critical curve.  Each point on the critical curve, then, corresponds to precise values of $N_f$, $N_b$, and $M$.  We show alternative versions of the critical curve in Fig.\ \ref{fig:M Nb}.  We note that the values of $N_f$ in Fig.\ \ref{fig:M Nb}(b) depend upon the arbitrarily chosen proportionality constant in Eq.\ (\ref{nf eos}).

As mentioned, we are strongly perturbing stable static solutions.  After introducing the perturbation, we compute $N_f$, $N_b$, and $M$ for the initial data.  We then determine the critical masses for the computed values of $N_f$ and $N_b$ using the curves in Fig. \ref{fig:M Nb}.  If $M$ is smaller than both critical masses, the system will move to the stable region.  If $M$ is not smaller than both critical masses, the system can either move to the stable region or collapse.  In this latter case, whether the system moves or collapses depends on the specifics of the perturbation (i.e.\ on the parameters $A$, $b$, $c$ in Eq.\ (\ref{perturbation})).  Determining the specifics is beyond the scope of this paper.

\begin{figure}
\centering
\includegraphics[width=3in]{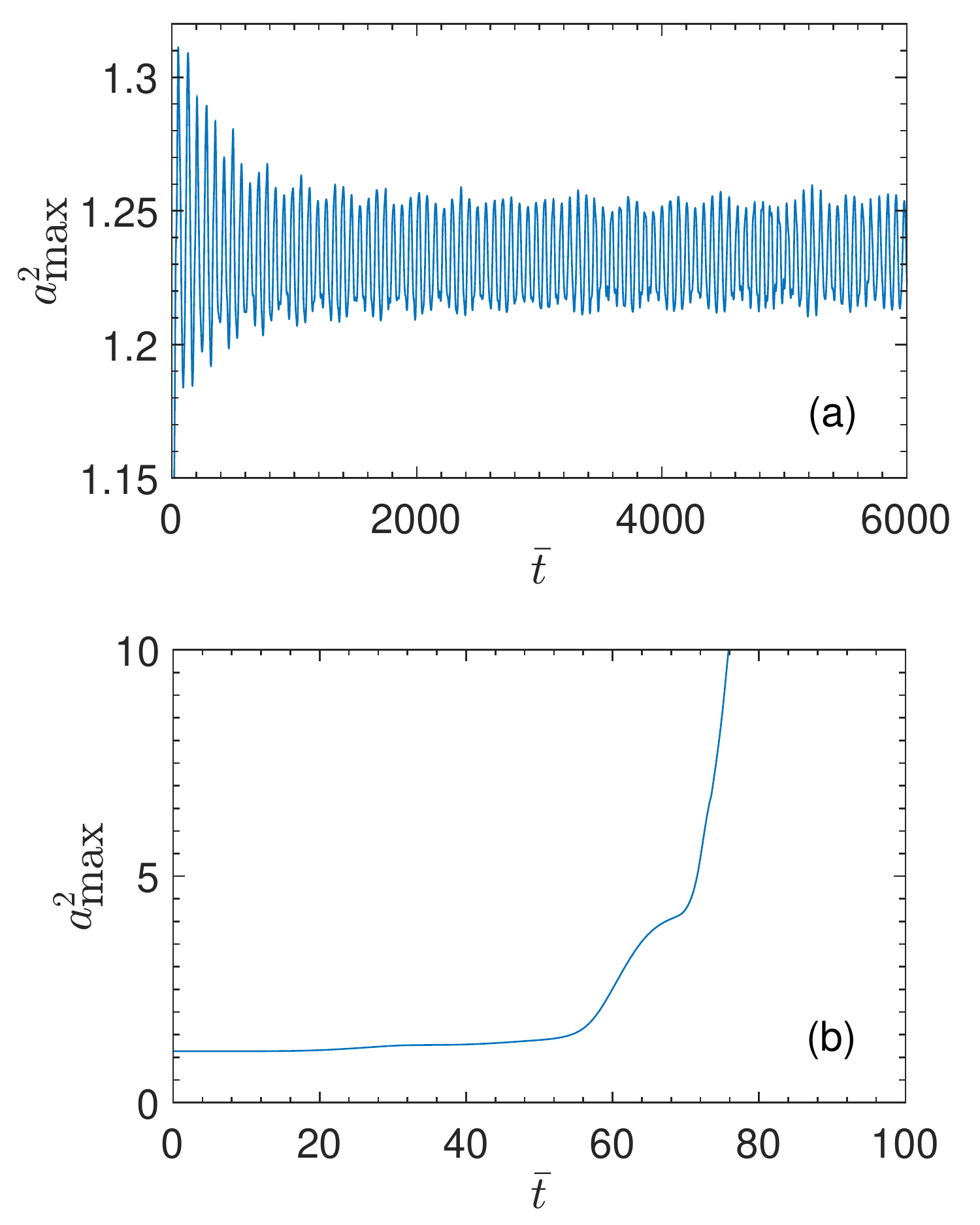}
\caption{Dynamical evolutions of a strongly perturbed stable static solution.  Both plots are for the same stable static solution, but are for different perturbations.  (a) The system migrates to a stable configuration.  (b) The system collapses to a black hole, as indicated by $a^2_\text{max} \rightarrow \infty$.}
\label{fig:dyn strong per}
\end{figure}

We show an example of this in Fig.\ \ref{fig:dyn strong per}.  Both plots begin with the same stable static solution, before perturbing, defined by $P(0) = 10^{1.5}$ MeV/fm$^3$ and $\varphi(0)/m_P = 10^{-1.5}$.  This static solution corresponds to the upper left blue dot in Fig.\ \ref{fig:dynamic stable}(a) and has mass $M = 0.733$ M$_\odot$.  In Fig.\ \ref{fig:dyn strong per}(a), the perturbed initial data have $M = 0.972$ M$_\odot$, $N_f = 4.127$, and $N_b = 0.507\times 10^{76}$.  After computing the critical masses for these values of $N_f$ and $N_b$ using Fig.\ \ref{fig:M Nb}, we find $M/M_\text{crit}(N_f) = 0.991$ and $M/M_\text{crit}(N_b) = 0.884$.  Since both of these ratios are less than 1, we expect the system to migrate to the stable region.  This is precisely what the periodic oscillations in Fig.\ \ref{fig:dyn strong per}(a) indicate.  In Fig.\ \ref{fig:dyn strong per}(b), the perturbed initial data have $M=1.038$ M$_\odot$, $N_f = 4.658$, and $N_b = 0.507\times 10^{76}$.  Again using Fig.\ \ref{fig:M Nb}, we find $M/M_\text{crit}(N_f) = 1.011$ and $M/M_\text{crit}(N_b) = 0.945$.  Since one of these ratios is greater than one, it is possible for the system to collapse to a black hole.  Indeed, Fig.\ \ref{fig:dyn strong per}(b) indicates precisely this since $a^2_\text{max} \rightarrow \infty$.

\section{Conclusion}
\label{sec:conclusion}

We presented dynamical evolutions of fermion-boson stars using a realistic equation of state for the ordinary nuclear matter.  We made a detailed study of the evolution of stable and unstable static solutions subject to weak and strong perturbations.  We found that weakly perturbed linearly stable static solutions can also be nonlinearly stable.  For weakly perturbed unstable static solutions, we found that the solution either migrates to a stable configuration or collapses to a black hole.  For strongly perturbed static solutions, such that the perturbation increases the total mass of the system, we found that unstable static solutions collapse and we identified the criteria under which stable static solutions move to the stable region and do not collapse.


\acknowledgments

We thank T.\ Gleason and B.\ Brown for help during the early stages of this work.  J.\ E.\ N.\ was funded by the 2021 Weiss Summer Research Program.



%

\end{document}